\documentclass[slac_one]{revtex4}
\usepackage{graphicx}
\usepackage{fancyhdr}
\RequirePackage{xspace}

\usepackage{relsize}
\def\babar{\mbox{\slshape B\kern-0.1em{\smaller A}\kern-0.1em
    B\kern-0.1em{\smaller A\kern-0.2em R}}}

\def\ccbar {\ensuremath{c\overline c}\xspace}
\def\piz   {\ensuremath{\pi^0}\xspace}

\def\Kbar  {\kern 0.2em\overline{\kern -0.2em K}{}\xspace}
\def\Kz    {\ensuremath{K^0}\xspace}

\def\KS    {\ensuremath{K^0_{\scriptscriptstyle S}}\xspace}
\def\Dbar    {\kern 0.2em\overline{\kern -0.2em D}{}\xspace}

\def\Dp      {\ensuremath{D^+}\xspace}
\def\Dm      {\ensuremath{D^-}\xspace}

\def\Dmp     {\ensuremath{D^\mp}\xspace}
\def\DpDm    {\ensuremath{\Dp {\kern -0.16em \Dm}}\xspace}
\def\Dstar   {\ensuremath{D^*}\xspace}

\def\Dstarp  {\ensuremath{D^{*+}}\xspace}
\def\Dstarm  {\ensuremath{D^{*-}}\xspace}
\def\Dstarpm {\ensuremath{D^{*\pm}}\xspace}

\def\B       {\ensuremath{B}\xspace}
\def\Bbar    {\kern 0.18em\overline{\kern -0.18em B}{}\xspace}

\def\BB      {\ensuremath{B\Bbar}\xspace}
\def\Bz      {\ensuremath{B^0}\xspace}
\def\Bzb     {\ensuremath{\Bbar^0}\xspace}
\def\BzBzb   {\ensuremath{\Bz {\kern -0.16em \Bzb}}\xspace}

\def\jpsi     {\ensuremath{{J\mskip -3mu/\mskip -2mu\psi\mskip 2mu}}\xspace}
\def\psitwos  {\ensuremath{\psi{(2S)}}\xspace}
\def\chicone  {\ensuremath{\chi_{c1}}\xspace}
\def\etac     {\ensuremath{\eta_c}\xspace}
\def\KL    {\ensuremath{K^0_{\scriptscriptstyle L}}\xspace}

\def\ks{\ensuremath{K^0_S}}
\def\kl{\ensuremath{K^0_L}}
\def\kstzero{\ensuremath{K^{*0}}}

\def\Y#1S{\ensuremath{\Upsilon{(#1S)}}\xspace}
\def\FourS {\Y4S}

\newcommand{\gev}{\ensuremath{\mathrm{\,Ge\kern -0.1em V}}\xspace}
\newcommand{\mev}{\ensuremath{\mathrm{\,Me\kern -0.1em V}}\xspace}
\newcommand{\gevc}{\ensuremath{{\mathrm{\,Ge\kern -0.1em V\!/}c}}\xspace}
\newcommand{\mevc}{\ensuremath{{\mathrm{\,Me\kern -0.1em V\!/}c}}\xspace}
\newcommand{\gevcc}{\ensuremath{{\mathrm{\,Ge\kern -0.1em V\!/}c^2}}\xspace}
\newcommand{\mevcc}{\ensuremath{{\mathrm{\,Me\kern -0.1em V\!/}c^2}}\xspace}
\def\to                 {\ensuremath{\rightarrow}\xspace}
\newcommand{\stat}{\ensuremath{\mathrm{(stat)}}\xspace}
\newcommand{\syst}{\ensuremath{\mathrm{(syst)}}\xspace}
\def\pep2{PEP-II}
\def\BF{$B$ Factory}

\def\CP                {\ensuremath{C\!P}\xspace}
\def\stwob{\ensuremath{\sin\! 2 \beta   }\xspace}

\def\deltat{\ensuremath{{\rm \Delta}t}\xspace}
\def\deltamd{\ensuremath{{\rm \Delta}m_d}\xspace}
\def\mistag{\ensuremath{w}\xspace}
\def\Imlambda{\ensuremath{\mathop{\cal I\mkern -2.0mu\mit m}\lambda}}
\def\abslambda{\ensuremath{|\lambda|}\xspace}
\def\Gammad{\ensuremath{\Gamma_d}\xspace}
\def\deltaGammad{\ensuremath{\Delta \Gammad}\xspace}

\def\jpsiks{\ensuremath{\jpsi\ks}}
\def\jpsikl{\ensuremath{\jpsi\kl}}

\def\psitwosks{\ensuremath{\psitwos\ks}}
\def\etacks{\ensuremath{\etac\ks}}
\def\chiconeks{\ensuremath{\chicone\ks}}
\def\jpsikstzero{\ensuremath{\jpsi\kstzero}}

\newcommand{\progtp}    [1]  {{Prog.\ Theor.\ Phys.\ {\bf #1}}}

\begin{document}

\title{\CP\ Violation in $\Bz$ decays to Charmonium and Charm Final States
at \babar } 

\author{Chunhui Chen}
\affiliation{Department of Physics, University of Maryland, 
College Park, Maryland 20742-4111, U.S.A \\Representing the \babar\ Collaboration}

\begin{abstract}
We report on measurements of time-dependent \CP-violation asymmetries in 
neutral \B\ meson decays to charmonium and charm final states. The results are obtained 
from a data sample of $(467\pm 5)\times 10^6$ 
$\FourS \to \BB$ decays collected with the \babar\ detector
at the PEP-II \B\ factory.  

\end{abstract}

\maketitle

\thispagestyle{fancy}

\section{Introduction}
In the Standard Model (SM), \CP violation is described by the
Ca\-bib\-bo-Ko\-ba\-ya\-shi-Mas\-ka\-wa (CKM) quark mixing matrix,
$V$~\cite{CKM}.  In particular an
irreducible complex phase in the $3\times3$ mixing matrix is the source of 
all the SM \CP violation.  Measurements of time-dependent \CP\ asymmetries
in the \Bz\ meson decays, through the interference between decays with 
and without $\Bz-\Bzb$ mixing, have provides stringent test on the mechanism 
of \CP\ violation in the SM. 

In this paper, we present
the most updated measurements of \CP\ violation in neutral \B\ meson decays
to charmonium and charm final states at \babar\ .  
The data used in this analysis were collected with the \babar\
detector operating at the \pep2\ \BF\ located at
the Stanford Linear Accelerator Center.  The \babar\ dataset comprises
$(467\pm 5)\times 10^6$ \BB pairs collected from 1999 to 2007 at the
center-of-mass (CM) energy $\sqrt{s}=10.58 \gev$, corresponding to the 
\FourS resonance.

\section{Analysis Technique}
To measure time-dependent \CP\ asymmetries, we typically fully reconstruct
a neutral \B\ meson decaying to a common final state ($B_{\mbox{rec}}$).
We identify (tag) the initial flavor of the $B_{\mbox{rec}}$
candidate using information from the other \B meson ($B_{\mbox{tag}}$) 
in the event. The decay rate $g_+$ $(g_-)$ for a neutral \B meson decaying 
to a \CP\ eigenstate accompanied by a \Bz (\Bzb) tag can be expressed as
\begin{equation}
g_\pm(\deltat) = \frac{e^{{- \left| \deltat \right|}/\tau_{\Bz} }}{4\tau_{\Bz} }
\Bigg\{ (1\mp\Delta\mistag) \Bigg.  \pm  (1-2\mistag)
\times \Big [ -\eta_f S\sin(\deltamd\deltat) -
 \Bigg. C\cos(\deltamd\deltat)  \Big] \Bigg\}\:\:\:
\label{eq:timedist}
\end{equation}
where 
\begin{equation}
S = -\eta_f \frac{2\Imlambda}{1+\abslambda^2},\;\;\;
C = \frac{1 - \abslambda^2 } {1 + \abslambda^2},\nonumber
\end{equation}
the \CP eigenvalue $\eta_f=+1$ ($-1$) for a \CP even (odd) final state,
$\deltat \equiv t_\mathrm{rec} - t_\mathrm{tag}$ is the difference
between the proper decay times of $B_{\mbox{rec}}$  and $B_{\mbox{tag}}$,
$\tau_{\Bz}$ is the neutral \B lifetime, and \deltamd is the mass difference
of the \B meson mass eigenstates
determined from $\Bz$-$\Bzb$ oscillations~\cite{Yao:2006px}.
We assume that
the corresponding decay-width difference \deltaGammad is zero.
Here, $\lambda=(q/p)(\bar{A}/A)$,
where $q$ and $p$ are complex constants that relate the \B-meson flavor
eigenstates to the mass eigenstates, and $\bar{A}/A$ is the ratio of
amplitudes of the decay without mixing of a \Bzb or \Bz to
the final state under study.
The average mistag probability \mistag describes the effect of incorrect
tags, and $\Delta\mistag$ is the difference between the mistag probabilities
for \Bz and \Bzb\ mesons.
The sine term in Eq.~\ref{eq:timedist} results from the interference
between direct decay and decay after $\Bz-\Bzb$ oscillation. A
non-zero cosine term arises from the interference between decay amplitudes
with different weak and strong phases (direct \CP violation $|\bar{A}/A|\neq 1$)
or from \CP
violation in $\Bz-\Bzb$ mixing ($|q/p|\neq 1$).

In the SM, \CP\ violation in mixing is negligible. When only one
diagram contributes to the decay process and no other weak phase 
appears in the 
process, we expect $C=0$ and $S=-\eta_f\stwob$ for \Bz\ decay that is
governed by a $b\to c$ transition, where $\beta\equiv
\text{arg}\left[-V_{cd}V_{cb}^*/V_{td}V_{tb}^*\right]$.

\section{\boldmath $\Bz\to(c\bar{c})K^{(*)0}$}
In the SM, the decay $\Bz\to(c\bar{c})K^{(*)0}$ is dominated by a
color-suppressed $b\to\ccbar s$ tree diagram and the
dominate penguin diagram has the same weak phase. As a result, 
the parameter $C=0$ and $S= -\eta_f\stwob$ are valid to a good
accuracy. Recent theoretical calculations suggest that the correction
on $S$ is of the order of $10^{-3}-10^{-4}$~\cite{Boos:2004xp}, well below the precision of
the current experimental measurement.

\begin{figure*}[htb]
\centering
\includegraphics[width=85mm]{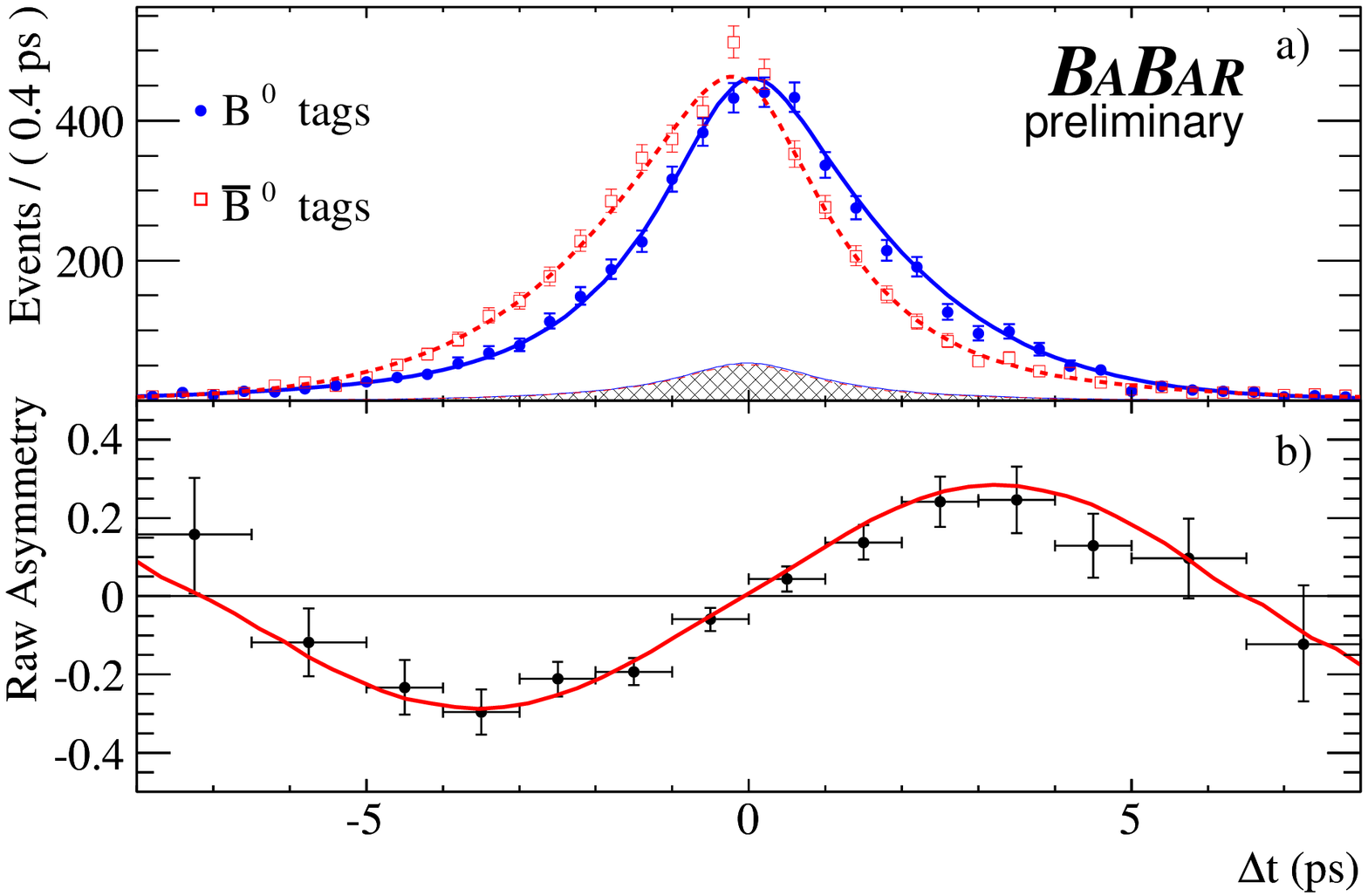}
\includegraphics[width=85mm]{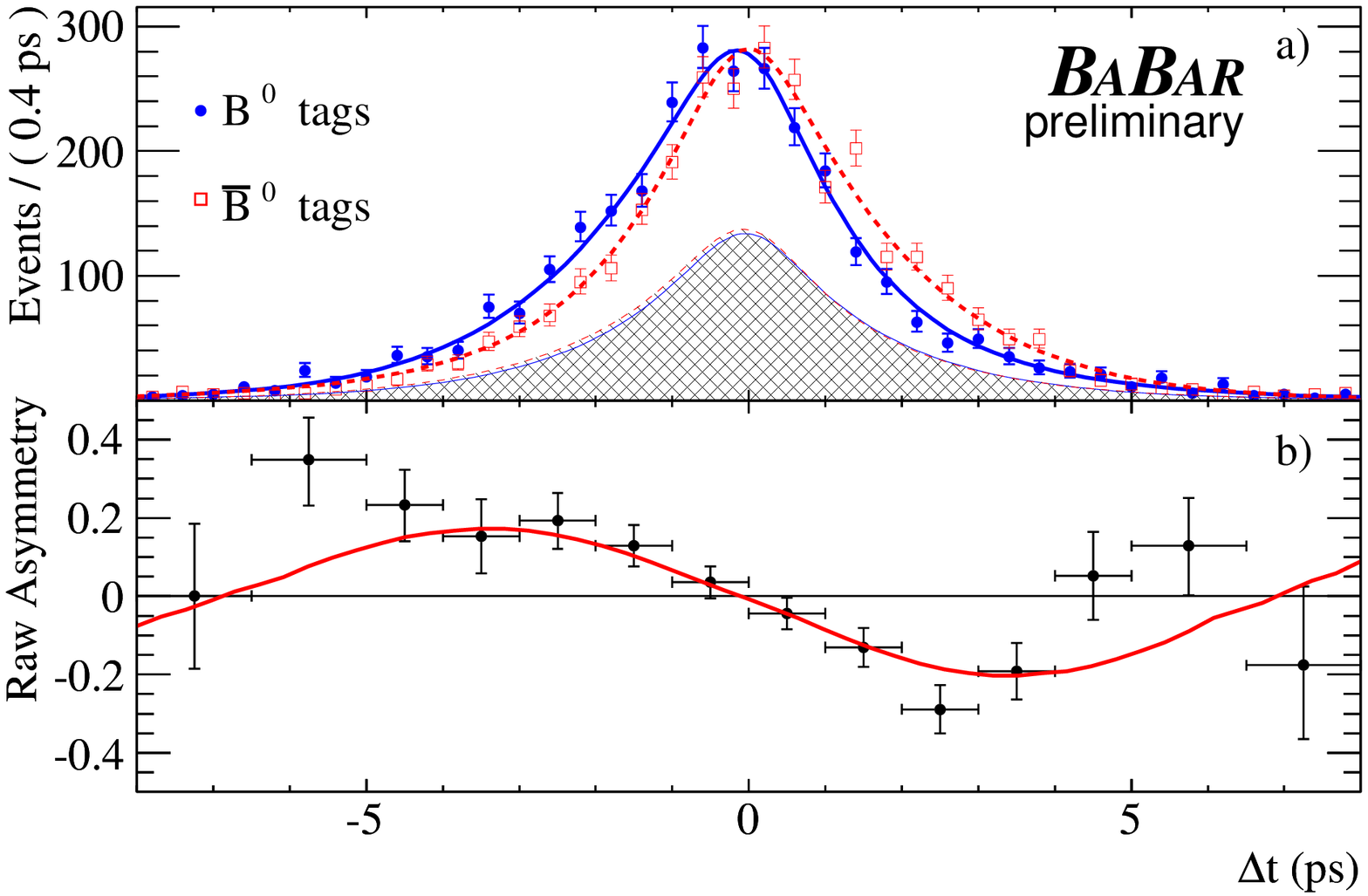}
\caption{Left: 
a) Number of $\eta_f=-1$ candidates ($\jpsi\KS$, $\psitwos\KS$,
$\chicone \KS$, and $\etac \KS$) in the signal region with a \Bz tag
($N_{\Bz }$) and with a \Bzb tag ($N_{\Bzb}$), and
b) the raw asymmetry, $(N_{\Bz}-N_{\Bzb})/(N_{\Bz}+N_{\Bzb})$, as functions
of \deltat;
Right: the corresponding distributions for the $\eta_f=+1$
mode $\jpsi\KL$.
The solid (dashed) curves represent the fit projections in \deltat for \Bz
(\Bzb) tags. The shaded regions represent the estimated background
contributions.
} 
\label{fig::stwob}
\end{figure*}

At \babar\ , we  reconstruct \Bz decays to the final states
\jpsiks, \jpsikl, \psitwosks, \chiconeks, \etacks, and
\jpsikstzero\ ($K^{(*)0}\to \KS \piz$).  
The $\jpsikl$ final state is
\CP even, and the $\jpsikstzero$ final state is an admixture of \CP even
and \CP odd amplitudes. Ignoring the angular information in $\jpsikstzero$ 
results in a dilution of the measured \CP asymmetry by a factor $1-2R_{\perp}$, 
where $R_{\perp}$ is the fraction of the $L$=1 contribution.
In Ref.~\cite{ref:rperp} we have measured 
$R_{\perp} = 0.233\pm\,\stat \pm 0.005\,\syst $,
which gives an effective $\eta_f = 0.504\pm 0.033 $ for $f=\jpsikstzero$,
after acceptance corrections. 
By assuming the same \CP\ asymmetries
for all the final states, we measure~\cite{Aubert:2008cp}
\begin{eqnarray}
S_f & = & 0.691 \pm 0.029\,\stat \pm 0.014\,\syst, \nonumber \\
C_f & = & 0.026 \pm 0.020\,\stat \pm 0.016\,\syst, \nonumber
\end{eqnarray}
where we define $C_f=\eta_f C$ and $S_f = \eta_f S$ to be consistent
with other time-dependent \CP asymmetry measurements.
In addition, \babar\ also performs measurements, including systematic 
uncertainties, using individual mode, because the theoretical
corrections could in principle be different among those modes.
The complete results can be found in reference~\cite{Aubert:2008cp}.
Our result is consistent with the SM expectation that $C_f=0$ and $S_f=\stwob$,
as well as our previous measurement~\cite{Aubert:2007hm}.
Figure~\ref{fig::stwob} shows the \deltat distributions and asymmetries 
in yields between \Bz tags and \Bzb tags for the
$\eta_f=-1$ and $\eta_f = +1$ samples as a function of \deltat, overlaid with 
the projection of the likelihood fit result. 

\section{\boldmath $\Bz\to D^{(*)\pm}D^{(*)\mp}$ }
The $\Bz\to D^{(*)\pm}D^{(*)\mp}$ decays are dominated by 
a  color-allowed tree-level $b\to\ccbar d$ transition. 
When neglecting the penguin (loop) amplitude, 
the mixing-induced \CP asymmetry of
$\Bz\to D^{(*)\pm}D^{(*)\mp}$ is also determined by 
\stwob~\cite{Sanda:1996pm}.  The effect of this assumption has
been estimated in models based on factorization and
heavy quark symmetry, and the corrections are expected to be a few
percent \cite{Xing:1998ca,Xing:1999yx}.  A large deviation of $S$
in $\Bz\to D^{(*)\pm}D^{(*)\mp}$ with respect to
\stwob determined from $b\to(\ccbar)s$ transition could indicate physics beyond the
SM~\cite{Grossman:1996ke,Gronau:2008ed,Zwicky:2007vv}.

The final state $\Dp\Dm$ is a \CP\ eigenstate so $S= -\stwob$ and $C=0$
in the SM when neglecting the penguin contribution. Most recently,  the 
Belle collaboration reported 
evidence of  large direct \CP\ violation in $\Bz\to\Dp\Dm$
where $C_{\Dp\Dm}=-0.91\pm 0.23\,\stat\pm 0.06\,\syst$~\cite{Fratina:2007zk}, 
in contradiction to the SM expectation.  
This has not been observed by \babar\ nor is it verified by the 
other $\Bz\to D^{*\pm}D^{(*)\mp}$ decay 
modes~\cite{Aubert:2007rr,Aubert:2007pa} which involve the
same quark-level diagrams. We updated our measurement with the complete
\FourS\ data sample collected at \babar\ and find~\cite{:2008aw}
\begin{eqnarray}
S_{\Dp\Dm} & = & -0.63 \pm 0.36\,\stat \pm 0.05\,\syst, \nonumber \\
C_{\Dp\Dm} & = & -0.07 \pm 0.23\,\stat \pm 0.03\,\syst, \nonumber
\end{eqnarray}
which is consistent with the SM with small penguin contributions.

The final state $D^{*\pm}D^{*\mp}$ is a mixture of \CP\ even and \CP\ odd 
states. The fraction of \CP\ odd component $R_T$ is determined to be
$R_T = 0.158 \pm 0.028\stat \pm 0.006\syst$ using a transversity analysis 
at \babar~\cite{:2008aw} . We performed a combined
analysis of the angular distribution and its time-dependence to extract
the time-dependent \CP\ asymmetry. We measure~\cite{:2008aw}
\begin{eqnarray}
S_+ &=& -0.76 \pm 0.16\,\stat \pm 0.04\,\syst, \nonumber \\
C_+ &=& \phantom{-}0.02 \pm 0.12\,\stat \pm 0.02\,\syst, \nonumber \\
S_\perp &=& -1.81 \pm 0.71\,\stat \pm 0.16\,\syst, \nonumber \\
C_\perp &=& \phantom{-}0.41 \pm 0.50\,\stat \pm 0.08\,\syst ,\nonumber
\label{eq:cptr}
\end{eqnarray}
where $S_+$ and $C_+$ are the \CP\ asymmetries of the \CP\ even
component, and $S_\perp$ and $C_\perp$ are the \CP\ asymmetries of 
the \CP\ odd component.
In the absence of penguin contribution, 
$S_+ = S_\perp = -\stwob$ and 
$C_+ = C_\perp = 0$.
Additionally we fit the data constraining $S_+ = S_\perp =
S_{\Dstarp\Dstarm}$ and $C_+ = C_\perp = C_{\Dstarp\Dstar}$ and 
find~\cite{:2008aw}
\begin{eqnarray}
S_{\Dstarp\Dstarm} &=& -0.71 \pm 0.16\,\stat \pm 0.03\,\syst \nonumber , \\
C_{\Dstarp\Dstarm} &=& \phantom{-}0.05 \pm 0.09\,\stat \pm 0.02\,\syst .\nonumber 
\label{eq:cprt1S}
\end{eqnarray}

Because $\Bz\to\Dstarpm\Dmp$ is not a \CP\ eigenstate, the expressions 
for $S$ and $C$ are related, 
$S_{\Dstarpm\Dmp} = -\sqrt{1 - C_{\Dstarpm\Dmp}}\sin(2\beta_{\mbox{eff}}\pm\delta)$, 
where $\delta$ is the strong phase difference between $\Bz\to\Dstarp\Dm$ and
$\Bz\to\Dp\Dstarm$~\cite{Aleksan:1993qk}.  When neglecting the penguin
contributions, $\beta_{\mbox{eff}}=\beta$, and 
$C_{\Dstarp\Dm}=-C_{\Dp\Dstarm}$. We measure~\cite{:2008aw}
\begin{eqnarray}
S_{\Dstarp\Dm} &= -0.62 \pm 0.21\,\stat \pm 0.03\,\syst ,\nonumber \\
S_{\Dp\Dstarm} &= -0.73 \pm 0.23\,\stat \pm 0.05\,\syst ,\nonumber \\
C_{\Dstarp\Dm} &= \phantom{-}0.08 \pm 0.17\,\stat\pm 0.04\,\syst ,\nonumber \\
C_{\Dm\Dstarp} &= \phantom{-}0.00 \pm 0.17\,\stat \pm 0.03\,\syst\,.\nonumber 
\end{eqnarray}

All of the $S$ parameters of $\Bz\to D^{(*)\pm}D^{(*)\mp}$ measured
in \babar\ are consistent with the value of \stwob\ measured in
$b\to\ccbar s$ transitions and with the expectation from the SM for small
penguin contributions. The $C$ parameters are consistent with zero
in all modes. In particular, we see no evidence of the large direct 
\CP\ violation reported by the Belle Collaboration in the 
$\Bz\to\Dp\Dm$ channel.

\section{\boldmath \Bz\to\jpsi\piz }
The $\Bz\to J/\psi \piz$ decay has the same quark level diagrams as $J/\psi\Kz$
except that the $s$ quark in $b\to\ccbar s$ is substituted by a $d$
quark. Therefore, the dominant tree diagram is Cabibbo suppressed compared to
that of $J/\psi\Kz$. However, unlike $J/\psi\Kz$, the dominant penguin diagram
in $J/\psi\piz$, whose CKM element factor is in the same order as the tree,
has a different weak phase from the tree. Therefore the deviation in $S$
from the \stwob\ of $J/\psi\Kz$ could be substantial.
This mode is also useful to constrain the penguin pollution in
$\Bz\to(\ccbar)\Kz$ mode in a more model-independent way (assuming
 SU(3) symmetry)~\cite{Ciuchini:2005mg}.

\babar\ has recently updated the \CP\ violation measurement in this
decay. We measure~\cite{Aubert:2008bs}
\begin{eqnarray}
S & = & -1.23 \pm 0.21\,\stat \pm 0.04\,\syst, \nonumber \\
C & = & -0.20 \pm 0.19\,\stat \pm 0.03\,\syst. \nonumber
\end{eqnarray}
The significance of $S$ or $C$ being nonzero has been estimated to be
$4.0\,\sigma$ using ensembles of MC simulated experiments including the
systematic uncertainties. This constitutes evidence of the
\CP\ violation in $\Bz\to J/\psi \piz$ decay. The numerical values of
$S$ and $C$ are consistent with the SM expectations
for a tree-dominated  $b\to\ccbar d$ transition.

\section{Conclusion}
We have presented most unadapted measurements of time-dependent 
\CP-violation asymmetries in 
neutral \B\ meson decays to charmonium and charm final states
from \babar\ experiment. The results are in a good agreement with the
SM expectations.


\begin{thebibliography}{99}

\bibitem{CKM}
\hyphenation{Ko-ba-ya-shi}
N.~Cabibbo, Phys.\ Rev.\ Lett.\ {\bf 10}, 531 (1963); M.~Kobayashi and T.~Maskawa, \progtp {\bf 49}, 652 (1973).

\bibitem{Yao:2006px}
W.~M.~Yao {\it et al.}  (Particle Data Group),
J.\ Phys.\ G {\bf 33} (2006) 1.

\bibitem{Boos:2004xp}
  H.~Boos, T.~Mannel and J.~Reuter,
  Phys.\ Rev.\  D {\bf 70}, 036006 (2004).

\bibitem{ref:rperp}
B.\ Aubert {\it et al.}, (\babar\ Collaboration), Phys.\ Rev.\ D {\bf 76}, 031102(R) (2007).

\bibitem{Aubert:2008cp}
B.~Aubert {\it et al.},  (\babar\ Collaboration),
arXiv:0808.1903 [hep-ex].

\bibitem{Aubert:2007hm}
B.~Aubert {\it et al.},  (\babar\ Collaboration),
  Phys.\ Rev.\ Lett.\  {\bf 99}, 171803 (2007).

\bibitem{Sanda:1996pm}
A.~I.~Sanda and Z.~z.~Xing,
Phys.\ Rev.\  D {\bf 56}, 341 (1997).

\bibitem{Xing:1998ca}
  Z.~z.~Xing,
  Phys.\ Lett.\  B {\bf 443}, 365 (1998).
\bibitem{Xing:1999yx}
Z.~z.~Xing,
  Phys.\ Rev.\  D {\bf 61}, 014010 (2000).

\bibitem{Grossman:1996ke}
  Y.~Grossman and M.~P.~Worah,
  Phys.\ Lett.\  B {\bf 395}, 241 (1997).

\bibitem{Gronau:2008ed}
  M.~Gronau, J.~L.~Rosner and D.~Pirjol,
  Phys.\ Rev.\  D {\bf 78}, 033011 (2008).

\bibitem{Zwicky:2007vv}
  R.~Zwicky,
  Phys.\ Rev.\  D {\bf 77}, 036004 (2008).

\bibitem{Fratina:2007zk}
  S.~Fratina {\it et al.}, (Belle Collaboration),
  Phys.\ Rev.\ Lett.\  {\bf 98}, 221802 (2007).

\bibitem{Aubert:2007rr}
B.~Aubert {\it et al.},  (\babar\ Collaboration),
  Phys.\ Rev.\  D {\bf 76}, 111102 (2007).

\bibitem{Aubert:2007pa}
B.~Aubert {\it et al.},  (\babar\ Collaboration),
  Phys.\ Rev.\ Lett.\  {\bf 99}, 071801 (2007).

\bibitem{:2008aw}
B.~Aubert {\it et al.},  (\babar\ Collaboration),
  arXiv:0808.1866 [hep-ex].

\bibitem{Aleksan:1993qk}
  R.~Aleksan, A.~Le Yaouanc, L.~Oliver, O.~Pene and J.~C.~Raynal,
  Phys.\ Lett.\  B {\bf 317}, 173 (1993).

\bibitem{Ciuchini:2005mg}
  M.~Ciuchini, M.~Pierini and L.~Silvestrini,
  Phys.\ Rev.\ Lett.\  {\bf 95}, 221804 (2005).

\bibitem{Aubert:2008bs}
B.~Aubert {\it et al.},  (\babar\ Collaboration),
  Phys.\ Rev.\ Lett.\  {\bf 101}, 021801 (2008).

\end{thebibliography}
\end{document}